\documentclass[twocolumn,showpacs,preprintnumbers,amsmath,amssymb]{revtex4-1}
\usepackage{graphicx}
\usepackage{dcolumn}
\usepackage{bm}

\usepackage[draft]{hyperref}
\usepackage{amsmath}
\usepackage{amssymb}

\begin{document}

\title{Deformable two-dimensional photonic crystal slab\\
for cavity optomechanics}

\author{Thomas Antoni,$^{*}$ Aur\'{e}lien G. Kuhn, Tristan Briant,
Pierre-Fran\c cois Cohadon, and Antoine Heidmann}

\address{Laboratoire Kastler Brossel, UPMC-ENS-CNRS, Case 74, 4 place
Jussieu, F75252 Paris Cedex 05, France\\
$^*$Corresponding author: thomas.antoni@spectro.jussieu.fr}

\author{R\'{e}my Braive, Alexios Beveratos, Izo Abram, Luc Le Gratiet, Isabelle Sagnes, and Isabelle Robert-Philip}

\address{Laboratoire de Photonique et Nanostructures LPN-CNRS, UPR-20,
Route de Nozay, 91460 Marcoussis, France}

\begin{abstract}
We have designed photonic crystal suspended membranes with
optimized optical and mechanical properties for cavity
optomechanics. Such resonators sustain vibration modes in the
megahertz range with quality factors of a few thousand. Thanks to
a two-dimensional square lattice of holes, their reflectivity at
normal incidence at 1064 nm reaches values as high as 95 $\%$.
These two features, combined with the very low mass of the
membrane, open the way to the use of such periodic structures as
deformable end-mirrors in Fabry-Perot cavities for the
investigation of cavity optomechanical effects.
\end{abstract}

%\ocis{120.4880, 160.4236, 160.5298, 230.5750, 220.4241.}

%]
\maketitle

\noindent Cavity optomechanics based on mechanically deformable
optical microcavities has attracted a lot of attention over the
past few years in view of applications in sensing and monitoring
of mechanical motion. A fundamental motivation behind such work
resides in the possibility of optically cooling the mechanical
motion, in view of reaching the quantum ground state of the
mechanical system \cite{Arcizet,PhysRevLett.97.243905}.
Downsizing the mechanical resonator allows
one to enhance the optomechanical coupling by reducing the
effective mass of the vibration modes. Among the various
nanomechanical oscillators presently investigated, photonic
crystal slabs may be good candidates, in particular due to the
versatility of their optical properties combined with their very
low mass. For instance, a ``defect'' in a planar one- or
two-dimensional periodic dielectric structure can offer a strong
confinement for both photons and phonons simultaneously inside a
very small volume, thus exhibiting a strong optomechanical
coupling \cite{Painter, TKIR}. The optomechanical response of two
coupled membranes, each pierced as a perfect two-dimensional
periodic dielectric structure sustaining band-edge Bloch modes,
has also been investigated \cite{PhysRevB.81.121101}.

In this paper, we present a new optomechanical device based on a
suspended photonic crystal membrane that can be used as a
deformable end-mirror in a Fabry-Perot cavity optomechanical
system. We investigate both its mechanical response (frequencies
and quality factors of the vibration modes) and its optical
response (reflectivity at normal incidence).

\begin{figure}[htb]
\centerline{\includegraphics[width=7.5cm]{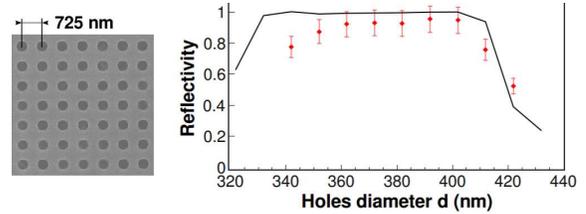}}
\caption{\label{optique}  Left: Scanning electron microscope
close-up view of the photonic crystal structure. Right: measured
(red dots) and simulated (black curve) reflectivity at $1064 \
\mathrm{nm}$ of a $10 \times 20\times 0.26  \ \mathrm{\mu m}^3$
fully clamped photonic crystal.}
\end{figure}

The photonic crystal consists of an InP slab pierced with a square
lattice of holes. The whole structure is grown by metal-organic
vapor phase epitaxy on an InP substrate, starting with a $1 \
\mathrm{\mu m}$ InGaAs sacrificial layer and then the InP slab
layer. Inductively coupled plasma etching of the photonic crystal
is performed on the top InP layer using a SiN hard mask defined by
electron-beam lithography and dry etching. To suspend the InP
membrane, the sacrificial layer under the membrane is removed by
selective wet etching \cite{talneau:061105}.

We were able to suspend membranes as large as $30 \times 30 \ {\mu
m}^2$, using several clamping solutions as described in the
following. We have optimized the photonic crystal structure in
order to maximize its optical reflectivity at normal incidence:
transmission through the slab and in-plane losses can be canceled
out by coupling the incident radiation to Bloch modes with a flat
dispersion curve at the $\Gamma$ point of the Brillouin zone.
Moreover, it is desirable to design a mirror with broadband
reflectivity so as to overcome fluctuations of the spectral
position of the high-reflectivity zone, which may be due to
defects and inhomogeneities introduced during the slab processing.
As suggested in \cite{Boutami:07}, this can be obtained by
coupling two non-degenerate but overlapping slow Bloch modes close
to the $\Gamma$ point with the radiated modes in the vertical
direction. Finite Difference Time Domain MEEP code
\cite{OskooiRo10} and eigenmode solver \cite{Johnson2001:mpb} MPB
have been used to determine adequate parameters for the photonic
crystal slabs in order to reach a near unity reflectivity over
more than 50 nm around 1064 nm. The simulations are performed
for an infinite structure, taking into account the evanescent
coupling and cavity effects with the underlying substrate.

\begin{figure}[htb]
\centerline{\includegraphics[width=8.5cm]{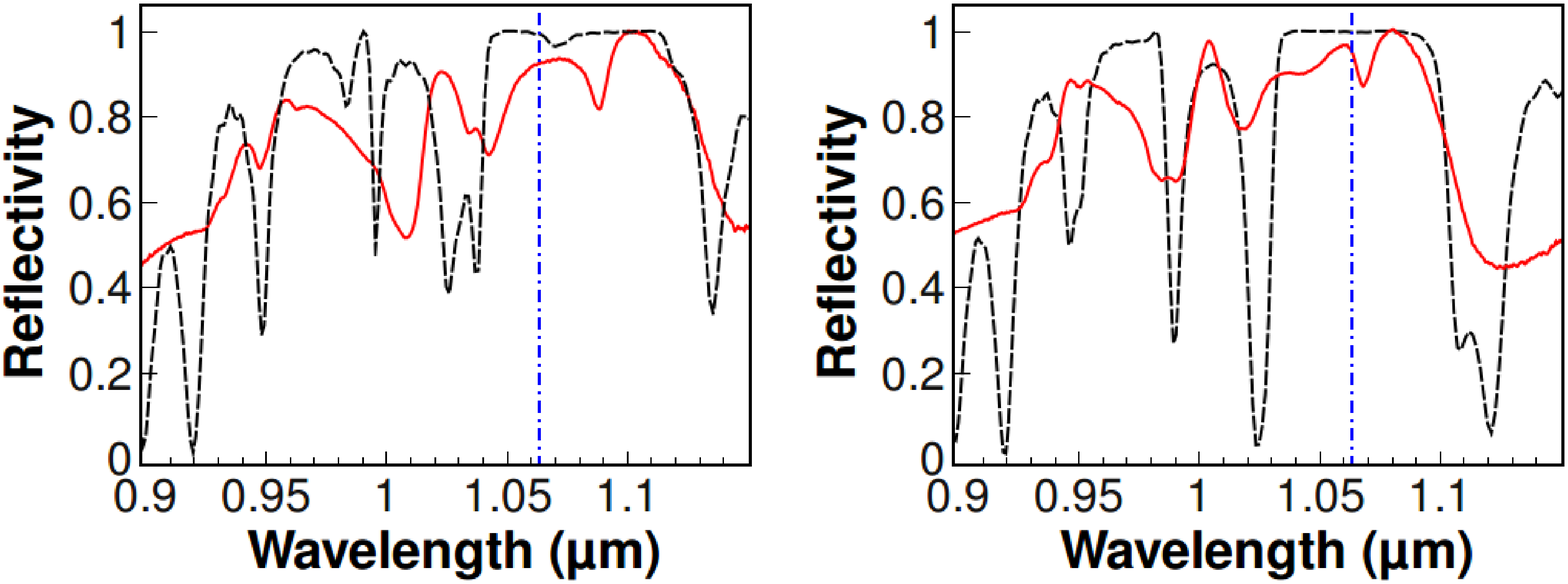}}
\caption{\label{spectres}Normalized experimental (red solid line)
and simulated (black dashed line) reflectivities as a function of
the wavelength, for a $30\times30\times 0.26\,\mathrm{\mu m}^3$
fully clamped photonic crystal membrane with holes diameters of
$372\ {\rm nm}$  (left) and $392\ {\rm nm}$ (right).}

\end{figure}

Fully clamped $30 \times 30 \ \mathrm{\mu m^2}$ samples with a
thickness of 260 nm have been processed to measure the slab's
reflectivity as a function of the optical wavelength by use of a
Fourier transform infrared spectroscopy setup. The corresponding
spectral behavior is shown in Figure \ref{spectres} for a photonic
crystal structure corresponding to a square array of holes, with a
period of 725 nm and different diameters (see Figure
\ref{optique}). The measured band edges well coincide with
reflectivity calculations, in particular the cut-off wavelengths
of the large plateau around $1064\ {\rm nm}$, and the resonance
about $1010\ {\rm nm}$. The spectral displacement of this
resonance with increasing hole diameters is also well reproduced.
The dip in the middle of the large plateau, mostly visible in the
calculated and measured spectra but with a depth that depends on
the hole diameter, is a consequence of the splitting between the
two underlying Bloch modes. Note that due to finite size effects
and optical diffraction, as well as to alignment and focusing
difficulties, the presented results are arbitrary normalized so
that absolute values of the curves are irrelevant. The reduced
visibility of the experimental reflectivity may be due as well to
finite size effects, as the beam waist is of the order of the
photonic crystal area, or to non-uniform hole diameters over the
membrane (resulting from proximity effects during the electron
beam lithography of such large photonic crystal surfaces
\cite{ieee}), or to the roughness of the substrate behind the
membrane, hence distorting the photonic crystal resonances.

Cavity operation of such membranes requires a very small beam
waist to minimize clipping loss. As a consequence the reflectivity
may be affected by the resulting beam divergence, about $6^\circ$
for a waist of $3 \ \mathrm{\mu m}$. To investigate this effect,
and also to measure an absolute value of the reflectivity,
we have used a small $10 \times 20 \ \mathrm{\mu m}^2$ fully
clamped membrane and focused a laser beam to a waist as small as
$2.5\ \mathrm{\mu m}$ with a microscope objective (0.65
numerical aperture). In addition, we have measured the dependence
of the reflectivity on the hole diameter. Experimental results
obtained by comparing the reflected power to the one reflected by
a $99.94 \%$ commercial mirror are shown in Figure \ref{optique}.
They are in good agreement with FDTD simulations, demonstrating
that divergence effects do not limit the reflectivity. The
absolute reflectivity reaches values as high as $95 \% \pm 8 \%$.

For the investigation of the mechanical response of the membrane,
we have built a Michelson-like interferometer in which the
membrane is the end-mirror in one of the arms. The light from a
$1064\,\mathrm{nm}$ Nd:YAG laser source is focused on the membrane
with the microscope objective, down to a beam waist of
$2.5\,\mathrm{\mu m}$. The membrane position is adjusted thanks
to a motorized 3-axis translation stage. The setup is inserted
into a vacuum chamber pumped down to $10^{-4}\, \mathrm{mbar}$.

\begin{figure}[htb]
\centerline{\includegraphics[width=8.5cm]{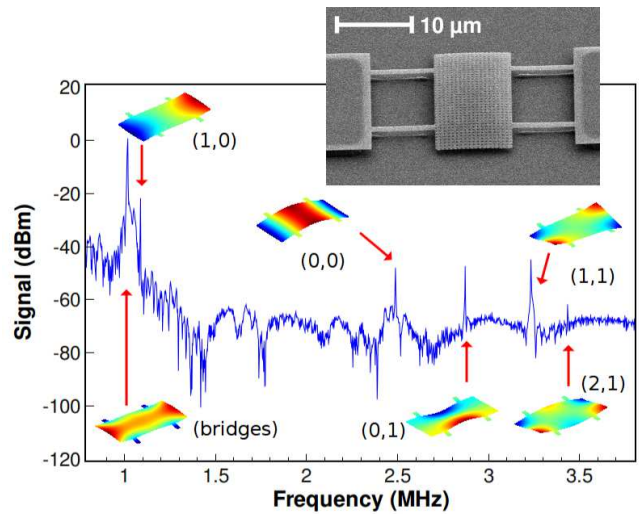}}
\caption{\label{mecanique} Displacement spectrum of a
$10\times20\times 0.21\,\mathrm{\mu m}^3$ photonic crystal
membrane, clamped by four $8\,\mu$m-long and $1\,\mu$m-wide
auxiliary bridges, as shown in the scanning electron microscope
image. The simulated vibration profile is shown for each
mechanical mode.}
\end{figure}

The interference between the two arms is monitored by a photodiode
and sent to a network analyzer which is also used to activate the
resonator via a piezoelectric stack. On the aforementioned
membranes, typical results show mechanical modes with resonance
frequencies in the megahertz range and with quality factors about
1000, due to clamping losses. We consequently have tried different
geometries and best results have been obtained with $10 \times 20
\ \mathrm{\mu m^2}$ suspended membranes, with a thickness of 210
nm. The membrane is suspended by four bridges with optimized
position, length and width (see inset in Fig. \ref{mecanique})
\cite{Aspelmeyer}: they are symmetrically located $5.9 \
\mathrm{\mu m}$ from the center of the membrane, at the node of
the fundamental longitudinal mode (0,0), as computed with
\textsc{Comsol Multiphysics}. We have tested samples with
auxiliary bridges length ranging from $2 \ \mathrm{\mu m}$ to $12
\ \mathrm{\mu m}$, and with a width down to $0.5 \ \mathrm{\mu
m}$.

The spectrum of the mechanical response obtained with a membrane
suspended by $8\,\mathrm{\mu m}$-long auxiliary bridges is shown
in Figure \ref{mecanique}. It displays six mechanical modes whose
frequencies $\Omega_\mathrm{m} / 2 \pi$ are in the megahertz
range, as expected from finite-element simulations, with the mode
of interest about $2.5\,{\rm MHz}$. The estimated effective mass
of this mode is of the order of $150 \ \mathrm{pg}$.

The quality factors $Q_\mathrm{m}$ of the mechanical modes are
inferred via a ring-down technique, by measuring the decay time of
the vibration mode after switching off the membrane actuation at
the resonance frequency. Quality factors ranging from $2\,000$ to
$10\,000$ are obtained, without relevant dependence on the
observed mode. We have investigated various dissipation mechanisms
that may limit these quality factors. For instance, we have
checked that $Q_\mathrm{m}$ does not depend on pressure below
$10^{-1} \ \mathrm{mbar}$, ruling out air viscous damping and the
well known squeezed-film damping\cite{0960-1317-14-10-003}. Using
Zener model \cite{PhysRevB.61.5600}, we have estimated that
thermoelastic damping is negligible as well. Another mechanism
that can be eliminated is the clamping losses corresponding to
dissipation into the substrate, since $Q_{\rm m}$ has no clear
dependence on the length or the position of the four auxiliary
bridges.

Two dominant dissipation channels can consequently be foreseen in
our system. The first one corresponds to surface effects
\cite{gaspar:622, liu:023524}, whose impact goes up for increasing
surface-to-volume ratio, estimated to $10\,\mathrm{\mu m^{-1}}$
(or 1 over 500 atoms) with our structures. Another loss mechanism
is related to crystal defects of the membrane, either induced in
the bulk material during processing, or due to a non-intentional
residual doping, estimated at about $10^{17}\,\mathrm{cm}^{-3}$
\cite{ekinci:061101, PhysRevB.66.085416}.

Note that the bending of the membrane distinguishable in Figure
~\ref{mecanique} is due to residual stress and yields to a radius
of curvature of about $100 \, \mathrm{\mu m}$. This is not an
issue for the design of a cavity since coupling mirrors with
smaller radius of curvature are achievable \cite{Reichel}.

Large optomechanical effects are expected by combining such
mechanical properties to high optical reflectivities, making
these photonic-crystal slabs promising candidates for cavity
optomechanics experiments. As an example, the shot-noise limited
displacement sensitivity with a single-ended Fabry-Perot cavity
would already be large enough to observe the Brownian and quantum
motions of the membrane. It is given by :
\begin{equation}
\label{shot} \delta x_{\rm shot} = \frac{\lambda}{16 \mathcal{F}
\sqrt{\overline{I}_{\rm in}}} \frac{T+L}{ T},
\end{equation}
where $\mathcal{F}$ is the cavity finesse, $T$ the transmission of
the coupling mirror and $L$ the cavity losses (including the
membrane reflectivity). Based on our experimental results, we find
$\delta x_{\rm shot} \simeq 2 \times 10^{-17}\
\mathrm{m/\sqrt{Hz}}$ for realistic parameters: incident power
$P_{\rm in}= \left(hc/\lambda \right) \overline{I}_{\rm in} = 5\
{\rm mW}$, $\mathcal{F} \simeq 100$, $L=5 \%$, and $T \simeq 1
\%$. This value is four orders of magnitude smaller than the Brownian
motion amplitude of the fundamental mode at 300 K ($\delta x_{T} =
\sqrt{2 Q_{\rm m} k_B T/(M \Omega_{\rm m}^3)} \simeq 2\times
10^{-13} \ \mathrm{m/\sqrt{Hz}}$), and of the same order of
magnitude as its zero-point motion ($\delta x_{T} \simeq 5 \times
10^{-17}\ \mathrm{m/\sqrt{Hz}}$ at $T\lesssim 150\,\mu$K).

Further improvement of the mechanical quality factors may allow
the experimental demonstration of the ground state of these
resonators, by performing cryogenic and laser coolings of the
membrane either used as an end-mirror in a Fabry-Perot cavity, or
as a high-reflectivity membrane in ``membrane-in-the-middle"
setups \cite{NatureJack}.

This research has been partially funded by the FP7 Specific Targeted
Research Projects QNems, and by the C'Nano Ile-de-France project
Naomi.

\end{document}